\pdfoutput=1
\documentclass[12pt]{iopart}
\usepackage[numbers,compress]{natbib}
\usepackage{xcolor}
\usepackage{graphicx}
\usepackage{booktabs}
\usepackage{lipsum}
\usepackage{subcaption}
\usepackage{diagbox}
\usepackage{relsize}
\usepackage[hidelinks]{hyperref}
\usepackage[noabbrev,nameinlink]{cleveref}
\usepackage{lineno}
\usepackage[separate-uncertainty=true]{siunitx}
\usepackage{threeparttable}
\usepackage{adjustbox}
\usepackage{float}
\usepackage[T1]{fontenc}
\usepackage{enumitem}
\usepackage{multirow}
\usepackage{lineno}

\crefformat{equation}{(#2#1#3)}
\def\newblock{\ }
\newcommand{\medianIQR}[3]{\qty{#1}~ \si{#3}~(\qty{#2}~ \si{#3})}

\begin{document}
\title[Analysis of Power Losses in Multichannel Electrical Stimulation Systems]{Analysis of Power Losses and the Efficacy of Power Minimization Strategies in Multichannel Electrical Stimulation Systems}
\author{F Varkevisser, W A Serdijn and T L Costa}
\address{Section Bioelectronics, Department of Microelectronics, Delft University of Technology, 
Mekelweg 4, 2628CD Delft, The Netherlands}
\ead{\{\href{mailto:f.varkevisser@tudelft.nl}{f.varkevisser};\href{mailto:w.a.serdijn@tudelft.nl}{w.a.serdijn};\href{mailto:t.m.l.dacosta@tudelft.nl}{t.m.l.dacosta}\}@tudelft.nl}
\vspace{10pt}
\begin{indented}
\item[]\today
\end{indented}

\begin{abstract}
\textit{Objective.} Neuroprosthetic devices require multichannel stimulator systems with an increasing number of channels. However, there are inherent power losses in typical multichannel stimulation circuits caused by mismatches between the power supply voltage and the voltage required at each electrode to successfully stimulate tissue. This imposes a bottleneck towards high-channel-count devices, which is particularly severe in wirelessly-powered devices. Hence, advances in the power efficiency of stimulation systems are critical. To support these advances, this paper presents a methodology to identify and quantify power losses associated with different power supply scaling  strategies in multichannel stimulation systems. 
\textit{Approach.} The methodology uses distributions of stimulation amplitudes and electrode impedances to calculate power losses in multichannel systems. Experimental data from prior studies spanning various stimulation applications were analyzed to evaluate the performance of fixed, global, and stepped supply scaling methods, focusing on their impact on power dissipation and efficiency. 
\textit{Main Results.} Variability in output conditions results in low power efficiency in multichannel stimulation systems across all applications. Stepped voltage scaling demonstrates substantial efficiency improvements, achieving an increase of \SIrange{43}{100}{\%}, particularly in high-channel-count applications with significant variability in tissue impedance. In contrast, global scaling proved effective only in systems with fewer channels and minimal inter-channel variation.
\textit{Significance.} The findings highlight the importance of tailoring power management strategies to specific applications to optimize efficiency while minimizing system complexity. The proposed methodology provides a framework for evaluating trade-offs between efficiency and system complexity, facilitating the design of more scalable and power-efficient neurostimulation systems.
\end{abstract}


\section{Introduction}
Implantable neurostimulation devices are widely used to treat neurological disorders such as Parkinson's disease, hearing loss, and visual impairment. Emerging applications, such as visual and bidirectional somatosensory prostheses, demand large-scale multichannel stimulator systems capable of stimulating hundreds to thousands of channels \cite{Fernandez2020TowardProspects, Musk2019AnChannels,Jung2024StableDevice}. The development of such systems is a complex interdisciplinary challenge, requiring intricate system- and circuit-level considerations for the electronic circuits \cite{Liu2020BidirectionalInterfaces}, and the design of biocompatible high-density electrode interfaces \cite{Drakopoulou2023HybridTools}. As the number of stimulation channels continues to scale, the available power becomes a major bottleneck. Traditionally, power is delivered wirelessly to the implantable stimulators since it avoids the infection risks posed by wired connections \cite{Nunen2023WirelessRectifier}. However, the power that can be transferred to the implant is limited by several safety regulations, such as the specific absorption rate (SAR) limit \cite{Nunen2023WirelessRectifier,ieee2005c95}. Consequently, optimizing the power efficiency of stimulator circuits is essential to enable further channel scaling and ensure these devices can function effectively within the limits of available power. 
Furthermore, power losses in the circuits lead to heat generation, which should be minimized to prevent damage to the tissue surrounding the implantable device \cite{ISO14708-3:2017}.
Improving power efficiency reduces excessive heating and improves the safety of the device.

\begin{figure}[tb]
    \centering
    \begin{subfigure}[b]{.31\columnwidth}
        \centering
        \includegraphics[width=\textwidth]{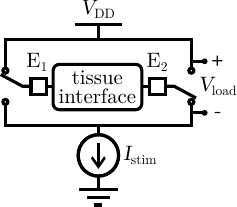}
        \caption{}
        \label{fig:driver}
   \end{subfigure}
   \hfill
   \begin{subfigure}[b]{.3\columnwidth}
        \centering
        \includegraphics[width=\textwidth]{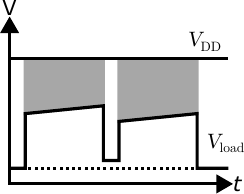}
        \caption{}
        \label{fig:Fixedlosses}
    \end{subfigure}
    \hfill
    \begin{subfigure}[b]{.3\columnwidth}
        \centering
        \includegraphics[width=\textwidth]{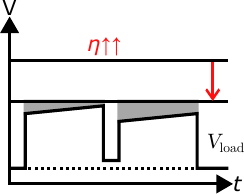}
        \caption{}
        \label{fig:ImprovedEfficiency}
    \end{subfigure}
    \caption{Illustration of the overhead losses in current-mode stimulation (CMS) in a bipolar electrode configuration. (a) Conventional output stage for CMS with a fixed voltage supply $V_{\textrm{DD}}$;  (b) Example of the load voltage ($V_{\textrm{load}}$) as a result of the current pulses delivered to the tissue. The mismatch between the load voltage and supply voltage (indicated in the grey area) leads to power dissipation in the output driver; (c) Illustration of how a scaled voltage supply can reduce the power dissipation in the output driver and thus increase the power efficiency, for the example in (b).}
    \label{fig:DriverLosses}
\end{figure}

Stimulator circuits are typically implemented to allow for current mode stimulation (CMS) or voltage mode stimulation (VMS). CMS is often preferred due to its precise control over injected charge, which is critical for safe stimulation \cite{Shirafkan2022Current-BasedTechniques, Merrill2005}. However, CMS suffers from inherent power inefficiency, as illustrated in \cref{fig:DriverLosses}. In a conventional bipolar CMS setup (\cref{fig:driver}), rectangular current pulses are generated from a fixed voltage supply, $V_{\textnormal{DD}}$. The stimulation current leads to a voltage drop over the tissue load equal to $V_{\textnormal{load}} = I_{\textnormal{stim}}Z_{\textnormal{tissue}}$ (\cref{fig:DriverLosses}), where $I_{\textnormal{stim}}$ is the stimulation current and $Z_{\textnormal{tissue}}$ the tissue impedance. Any mismatch between $V_{\textnormal{load}}$ and $V_{\textnormal{DD}}$ (indicated in grey in \cref{fig:DriverLosses}) leads to excessive power dissipation in the current source, reducing overall efficiency. A possible solution is to scale down the voltage supply, minimizing overhead losses and improving efficiency ($\eta$), as illustrated in \cref{fig:ImprovedEfficiency}. 

\begin{figure*}[tb]
    \centering
    \begin{subfigure}[b]{.45\columnwidth}
        \centering
        \includegraphics[width=\textwidth]{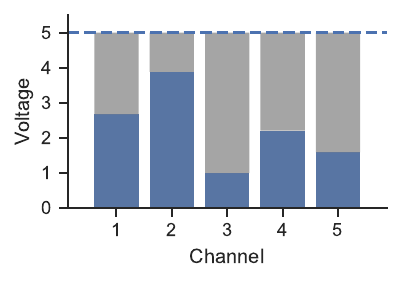}
        \caption{}
        \label{fig:FixedSupply}
    \end{subfigure}
    \hfill
    \begin{subfigure}[b]{.45\columnwidth}
        \centering
        \includegraphics[width=\textwidth]{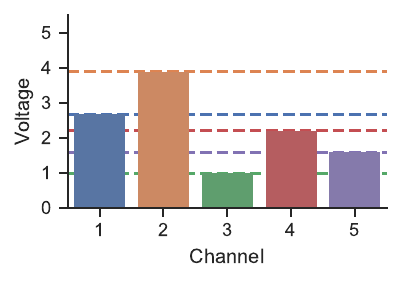}
        \caption{}
        \label{fig:IdealSupply}
   \end{subfigure}
    \begin{subfigure}[b]{.45\columnwidth}
        \centering
        \includegraphics[width=\textwidth]{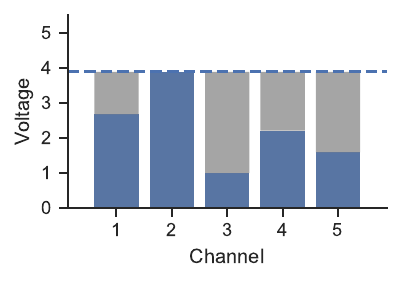}
        \caption{}
        \label{fig:GlobalScaling}
    \end{subfigure}
    \hfill
    \begin{subfigure}[b]{.45\columnwidth}
        \centering
        \includegraphics[width=\textwidth]{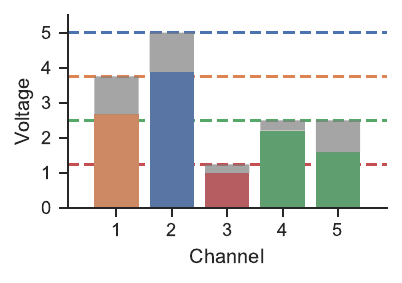}
        \caption{}
        \label{fig:SteppedSupply}
    \end{subfigure}
    \caption{Illustration of the overhead losses for different voltage scaling strategies in the example of a system with five channels. Dashed, horizontal lines indicate the available voltage rails, and the colored bars the load voltage ($V_{\textrm{load}}$) of the specific channel, where the color indicates to which voltage rail the channel is connected. Grey rectangles indicate the overhead losses. (a) In the case of a fixed voltage supply, all channels share the same voltage supply. (b) In the case of ideal supply scaling, each channel has a specific voltage supply matched to its load voltage. Thus, the overhead losses are zero. (c) With a global supply scaling strategy, a shared supply voltage is scaled to the worst-case $V_{\textrm{load}}$ (channel 2 in the example), eliminating all overhead losses for that channel and reducing overhead losses in the other channels compared to the fixed voltage strategy. (d) A stepped voltage supply strategy with \num{4} rails. Multiple voltage rails are available, and each channel is connected to the nearest rail above its load voltage. }
    \label{fig:ScalingStrategies}
\end{figure*}

In multichannel systems, however, the variability of tissue impedance and current amplitude between channels complicates this approach. Each channel has different voltage requirements, making the application of voltage scaling more complex \cite{Davis2012}. To address this, several voltage scaling strategies are proposed in the literature to reduce the losses at the output driver. The different scaling strategies are illustrated in \cref{fig:ScalingStrategies} for a system with five channels with varying load voltage requirements. 
\Cref{fig:FixedSupply} illustrates the conventional approach of using a fixed voltage supply for all channels. Ideally, each channel would have its own dedicated voltage supply precisely matching its load voltage (\cref{fig:IdealSupply}), a strategy often referred to as adiabatic voltage scaling \cite{Arfin2012,Kelly2022AdiabaticStimulator}. However, this approach faces scalability limitations, as each channel requires a separate voltage supply, and implementing adiabatic scaling often demands an impractically large chip area, making it unsuitable for scalable solutions.
An alternative is to use a single scalable supply (\cref{fig:GlobalScaling}) \cite{Luo2017,Williams2013}, where the supply voltage is configured to accommodate the worst-case channel (Channel 2 in \cref{fig:GlobalScaling}). While this ensures high efficiency for the worst-case channel, it can result in significant overhead losses for other channels.
Another strategy involves creating $N$ voltage rails distributed across all channels \cite{Eom2023AScaling,Nguyen2023AProcess,Rashidi2021FullyRecovery,Lee202427.3Devices}, with each channel connected to the nearest rail above its load voltage (\cref{fig:SteppedSupply}). This approach offers a trade-off between power efficiency and system complexity with the choice of $N$.

The impact of inter-channel variability on the power efficiency of multichannel stimulation systems is often neglected in conventional designs. This work introduces a novel methodology that incorporates these effects to evaluate the efficacy of various supply scaling strategies and quantify the associated power losses. Using experimental data from various multichannel stimulation applications, the methodology calculates channel-specific load voltage requirements and estimates the overhead losses for different voltage scaling strategies. The results provide valuable insights into the trade-offs between power efficiency and design complexity, offering a systematic framework to guide circuit design considerations for large-scale neurostimulation systems.

\section{Methods}
\subsection{Data Collection and Extraction}\label{subsec:DataExtraction}
Experimental data was collected using a systematic search in the Scopus database for studies on (micro)stimulation. Inclusion criteria required that the selected studies report (perception) thresholds and impedance data and that the subjects are either humans or non-human primates. For some cases with partially available data, the authors were contacted to request additional data. We collected 26 datasets from 7 publications, spanning four applications. The results are organized by application, resulting in categories for intracortical visual prostheses (V1), retinal implants, intrafascicular peripheral nerve stimulation (iPNS), and extraneural PNS. The extracted datasets are detailed below. Each study has its own definitions and methods of collecting and reporting the data. All values are reported as (mean $\pm$ sd) unless stated otherwise. 

The study by Fernández et al. \cite{Fernandez2021VisualCortex} explored the use of a Utah Electrode Array (UEA) with \num{96} electrodes implanted in the visual cortex of a human patient. The authors applied stimulation trains of \num{50} monopolar charge-balanced cathodic-first biphasic
stimuli, with a pulse width (PW) of \SI{170}{\us}, an interphase delay (IPD) of \SI{60}{\us}, and a repetition frequency of \SI{300}{\Hz} for the threshold procedure. A binary search procedure was used to determine the stimulation threshold that led to a visual perception in \SI{50}{\%} of the trials. The magnitude of the electrode impedance was measured at \SI{1}{\kHz}. The current threshold for single-electrode stimulation was \SI{66.8 \pm 36.5}{\uA}, while electrode impedances of \SI{47 \pm 4.8}{\kohm} were recorded. 

Building on similar stimulation parameters, Chen et al. \cite{Chen2023ChronicMonkeys} implanted \num{1024} microelectrodes (\num{16} \num{64}-channel UEAs) in the visual cortex of two monkeys ('Monkey A' and 'Monkey L'). For both monkeys, two current thresholds are reported, one in the early stage after implantation ($\mu_{\textrm{early}}$) and one in the late stage of the experiments ($\mu_{\textrm{late}}$). The reported thresholds are $\mu_{\textrm{early}}$ = \SI{65 \pm 45}{\uA} and $\mu_{\textrm{late}}$ = \SI{60 \pm 58}{\uA} for Monkey A, and $\mu_{\textrm{early}}$ = \SI{19 \pm 17}{\uA} and $\mu_{\textrm{late}}$ = \SI{80 \pm 71}{\uA} for Monkey L. Additionally, we received the dataset of recorded electrode impedances from the authors. From this dataset, we obtained the impedance values (at \SI{1}{\kHz}) for both monkeys in the first and last experiment sessions. These values were filtered to include only electrodes with impedance below \SI{300}{\kohm} for Monkey A and below \SI{150}{\kohm} for Monkey L, as these electrodes were used for the current threshold measurements \cite{Chen2023ChronicMonkeys}. In Monkey A, the recorded electrode impedance was \SI{144.7 \pm 72.6}{\kohm} and \SI{71.1 \pm 70.6}{\kohm} in the early and late stages, respectively, and for Monkey L, it was \SI{75.1 \pm 36.3}{\kohm} and \SI{74.9 \pm 36.4}{\kohm}.

While Fernández and Chen focused on cortical implants, De Balthasar et al. \cite{DeBalthasar2008FactorsProstheses} investigated epiretinal implants in six human subjects (S1-S6). Since the impedance data is only reported for subjects S4-S6, the other subjects are not included in this analysis. The array consisted of 16 (4x4) platinum electrodes in a checkerboard arrangement with alternating electrode sizes of \SI{260}{\um} and \SI{520}{\um}. The stimulation thresholds were determined using single-pulse symmetric cathodic-first pulses with a PW of \SI{0.975}{\ms} and an IPD of \SI{0.975}{\ms}. The perceptual thresholds, in this case, are defined as the current amplitude that causes a percept in \SI{79}{\%} of the trials. The electrode impedance was reported separately for the two sizes and are therefore treated as separate datasets in this analysis. The reported current thresholds ($\mu_{subject,size}$) are: $\mu_{\textrm{S4,260}}$ = \SI{233 \pm 20.9}{\uA}, $\mu_{\textrm{S5,260}}$ = \SI{30.3 \pm 1.7}{\uA}, $\mu_{\textrm{S6,260}}$ = \SI{40.9 \pm 6.1}{\uA}, $\mu_{\textrm{S4,520}}$ = \SI{222.9 \pm 16}{\uA}, $\mu_{\textrm{S5,520}}$ = \SI{26.9 \pm 1.3}{\uA}, $\mu_{\textrm{S6,520}}$ = \SI{37.8 \pm 4.9}{\uA}. Furthermore, the reported impedances are as follows: $Z_{\textrm{S4, 260}}$ = \SI{25.6 \pm 3}{\kohm}, $Z_{\textrm{S5, 260}}$ = \SI{40.8 \pm 1.5}{\kohm}, $Z_{\textrm{S6, 260}}$ = \SI{36.5 \pm 1.8}{\kohm}, $Z_{\textrm{S4, 520}}$ = \SI{13.6 \pm 1.1}{\kohm}, $Z_{\textrm{S5, 520}}$ = \SI{22.9 \pm 0.3}{\kohm}, $Z_{\textrm{S6, 520}}$ = \SI{18.7 \pm 0.4}{\kohm}. 

Similarly targeting retinal stimulation, Demchinsky et al. \cite{Demchinsky2019TheWhy} implanted one human patient with the Argus II \cite{ArgusII} retinal prosthesis. The Argus II has an epiretinal electrode array of \num{6}$\times$\num{10} platinum electrodes with a diameter of \SI{200}{\um} \cite{ArgusII}. The parameters for stimulation thresholds and impedance measurements are not specified in this study, but the perception threshold is reported as the amplitude evoking a visual percept in \SI{50}{\%} of the trials. The measured perception threshold and electrode impedance after six months of implantation were \SI{251 \pm 197}{\uA} and \SI{5.10 \pm 1.77}{\kohm}\textsuperscript{\footnotemark}, respectively. 
\footnotetext{In \cite{Demchinsky2019TheWhy}, the current and impedance values are reported in \si{\mA} and \si{\ohm}, respectively in Table 2. Based on impedance and current values reported in other Argus II studies, it is assumed that this is a mistake and that these values should be \si{\uA} and \si{\kohm} instead.}

Instead of targeting the central nervous system, Tan et al. \cite{Tan2015StabilityAmputees} examined extraneural peripheral nerve stimulation (PNS) in two human amputees. They implanted flat interface nerve electrodes (FINE) around the median and ulnar nerves in their mid-forearm and Case Western Reserve University (CWRU) electrodes around the radial nerve to produce selective sensory responses. The FINE electrode around the ulnar nerve in subject 2 did not retain good contact with the nerve and is therefore not included in the results. The stimulation pattern was a pulse train (\SI{100}{\Hz}) of monopolar, bi-phasic, charge-balanced, cathodic-first square pulses, with a sinusoidal modulated pulse width (\SI{1}{\Hz}) to evoke a natural, pulsing perception. Furthermore, the PW, as well as the amplitude, were stepped during the threshold process. As a result, the stimulation charge is used to report the perception threshold.
However, the current amplitude is required for the power loss calculations presented in this work. To estimate the current amplitude from the reported charge thresholds, the average PW is estimated at \SI{100}{\us}, and the current threshold is calculated as $I_{\textrm{th},i} = Q_{\textrm{th},i}/100\mu s$. The resulting perception current thresholds are: $\mu_{\textrm{S1,median}}$ = \SI{0.96 \pm 0.43}{\mA}, $\mu_{\textrm{S1,ulnar}}$ = \SI{0.71 \pm 0.59}{\mA}, $\mu_{\textrm{S1,radial}}$ = \SI{0.41 \pm 0.12}{\mA}, $\mu_{\textrm{S2,median}}$ = \SI{1.26 \pm 0.42}{\mA}, $\mu_{\textrm{S2,radial}}$ = \SI{1.20 \pm 0.33}{\mA}. 
To measure the electrode impedance, \SI{0.3}{\mA} and \SI{50}{\us} pulses at \num{20} and \SI{100}{\Hz} between pairs of electrodes within each cuff were used. The mean of eight measures of the resulting peak voltage drop between each pair of contacts was measured to calculate the impedance. The reported impedances are: $Z_{\textrm{S1,median}}$ = \SI{3.12 \pm 0.15}{\kohm}, $Z_{\textrm{S1,ulnar}}$ = \SI{2.66 \pm 0.15}{\kohm}, $Z_{\textrm{S1,radial}}$ = \SI{2.91 \pm 0.22}{\kohm}, $Z_{\textrm{S2,median}}$ = \SI{2.92 \pm 0.21}{\kohm}, $Z_{\textrm{S2,radial}}$ = \SI{3.09 \pm 0.19}{\kohm}.

Finally, Davis et al. \cite{Davis2016RestoringNerves} and George et al. \cite{George2020Long-termMuscles} focused on intrafascicular PNS (iPNS) using Utah Slanted Electrode Arrays (USEA). In \cite{Davis2016RestoringNerves}, two human amputees were implanted with a USEA of \num{96} electrodes in the sensory nerves in the forearm. The USEA delivers intrafascicular microstimulation, in contrast to the electrodes used in \cite{Tan2015StabilityAmputees}, which wrap around the nerve. In subject \num{1}, the array was implanted in the median nerve, while in subject \num{2}, it was implanted in the ulnar nerve. Biphasic, cathodic-first stimulation was used to determine the perception threshold. In most thresholding experiments, a constant frequency of \SI{200}{\Hz} and train duration of \SI{0.2}{\s} were used. The resulting perception thresholds were $\mu_{\textrm{S1}}$ = \SI{27.0 \pm 20}{\uA}, $\mu_{\textrm{S2}}$ = \SI{12.0 \pm 11.0}{\uA}. 
The electrode impedance was measured using a sinusoidal current at \SI{1}{\kHz} through a reference electrode. Electrodes with an impedance \SI{<500}{\kohm} were defined as working electrodes. The measured impedances for the working electrodes are $Z_{\textrm{S1}}$ = \SI{222 \pm 133}{\kohm} and $Z_{\textrm{S2}}$ = \SI{143 \pm 76}{\kohm}. The number of working electrodes in subject \num{1} rapidly dropped over the duration of the study. 

A complementary study by George et al. \cite{George2020Long-termMuscles} provides further insight into the long-term viability of iPNS in humans. In \cite{George2020Long-termMuscles}, two human amputees (S5 \& S6) were chronically implanted with USEAs in their residual arm nerves to restore sensorimotor function. In both participants, one array was implanted in the median nerve and one in the ulnar nerve. The study included a third participant (S7). However, the stimulation thresholds are only reported for S5 \& S6. Therefore, S7 is excluded from the analysis in this work. Contrary to the other studies, the perception threshold values in \cite{George2020Long-termMuscles} are reported in the format 'median (IQR).' The perception thresholds are reported for the first and last session, leading to a total of 8 datasets; however, since the last session in S6-ulnar contains a very limited set of electrodes, it is left out of the analysis in this work. The reported perception thresholds are as follows: $\mu_{\textrm{S5M,first}}$ = \SI{25}{\uA}~(\SI{17}{\uA}), $\mu_{\textrm{S5U,first}}$ = \SI{31}{\uA}~(\SI{31}{\uA}), $\mu_{\textrm{S6M,first}}$ = \SI{21}{\uA}~(\SI{11}{\uA}), $\mu_{\textrm{S6U,first}}$ = \SI{36.5}{\uA}~(\SI{42.5}{\uA}), $\mu_{\textrm{S5M,last}}$ = \SI{60}{\uA}~(\SI{40}{\uA}), $\mu_{\textrm{S5U,last}}$ = \SI{70}{\uA}~(\SI{52.5}{\uA}), $\mu_{\textrm{S6M,last}}$ = \SI{72.5}{\uA}~(\SI{25}{\uA}). The electrode impedance data is only shown in figure format in the paper. The data behind this figure was provided to us by the authors and we used the impedance data of the first and last sessions of each participant for our analysis. The recorded impedance data is as follows: $Z_{\textrm{S5M,first}}$ = \SI{81.6}{\kohm}~(\SI{99.6}{\kohm}), $Z_{\textrm{S5U,first}}$ = \SI{77.5}{\kohm}~(\SI{101.5}{\kohm}), $Z_{\textrm{S6M,first}}$ = \SI{67.6}{\kohm}~(\SI{130.6}{\kohm}), $Z_{\textrm{S6U,first}}$ = \SI{49.0}{\kohm}~(\SI{71.4}{\kohm}), $Z_{\textrm{S5M,last}}$ = \SI{131.3}{\kohm}~(\SI{163.8}{\kohm}), $Z_{\textrm{S5U,last}}$ = \SI{178.6}{\kohm}~(\SI{65.9}{\kohm}), $Z_{\textrm{S6M,last}}$ = \SI{50.6}{\kohm}~(\SI{42.8}{\kohm}).

The datasets are summarized in \cref{tab:Datasets}.

\begin{table}[t]
    \centering
    \caption{Summary of the datasets used in this work. All numerical data is presented as 'mean $\pm$ sd,' except for \cite{George2020Long-termMuscles}, where it is presented as 'median (IQR).' U(S)EA = Utah (Slanted) Electrode Array, FINE = Flat Interface Nerve Electrode, V1 = Primary visual cortex, and (i)PNS = (intrafascicular) Peripheral Nerve Stimulation.}
    \label{tab:Datasets}
    \begin{adjustbox}{max width=\columnwidth}
    \begin{threeparttable}[b]
    \sisetup{separate-uncertainty=true, table-number-alignment=center, table-align-text-post = false}
    \begin{tabular}{@{}cccS[table-format = 3.1(4)]S[table-format = 4.0(3)]cc@{}}
    \toprule
    \# & Source & Dataset & {\shortstack{Electrode \\ impedance [k$\Omega$]}} & {\shortstack{Current \\ threshold [$\mu$A]}} & Target & Electrodes \\ \midrule
    1 & \cite{Fernandez2021VisualCortex} & Human & 47(4.8) & 67(37) &  V1 & UEA \\
    2 & \cite{Chen2023ChronicMonkeys} & Monkey A early & 144.7(72.6)  & 65(45) & V1 & UEA \\
    3 & \cite{Chen2023ChronicMonkeys} & Monkey A late   & 71.1(70.6) & 60(58) & V1 & UEA\\
    4 & \cite{Chen2023ChronicMonkeys} & Monkey L early & 75.1(36.3) & 19(17) & V1 & UEA\\
    5 & \cite{Chen2023ChronicMonkeys} & Monkey L late & 74.9(36.4) & 80(71) & V1 & UEA\\ 
    6 & \cite{DeBalthasar2008FactorsProstheses} & S4 260$\mu$m & 25.6(3) & 233(21) & Retina & Custom \\
    7 & \cite{DeBalthasar2008FactorsProstheses} & S5 260$\mu$m & 40.8(1.5) & 30(2) & Retina & Custom \\
    8 & \cite{DeBalthasar2008FactorsProstheses} & S6 260$\mu$m & 36.5(1.8) & 41(6) & Retina & Custom \\
    9 & \cite{DeBalthasar2008FactorsProstheses} & S4 520$\mu$m & 13.6(1.1) & 222(16) & Retina & Custom \\
    10 & \cite{DeBalthasar2008FactorsProstheses} & S5 520$\mu$m & 22.9(0.3) & 27(1) & Retina & Custom \\
    11 & \cite{DeBalthasar2008FactorsProstheses} & S6 520$\mu$m & 18.7(0.4) & 38(5) & Retina & Custom \\
    12 & \cite{Demchinsky2019TheWhy} & Human & 5.1(1.8) & 251(197) & Retina & Argus II \\
    13 & \cite{Tan2015StabilityAmputees} & S1 median & 3.1(0.2) & 955(425) & PNS & FINE \\
    14 & \cite{Tan2015StabilityAmputees} & S1 ulnar & 2.7(0.2) & 707(592) & PNS & FINE \\
    15 & \cite{Tan2015StabilityAmputees} & S1 radial & 2.9(0.2) & 407(124) & PNS & FINE \\
    16 & \cite{Tan2015StabilityAmputees} & S2 median & 2.9(0.2) & 1260(415) & PNS & FINE \\
    17 & \cite{Tan2015StabilityAmputees} & S1 radial & 3.1(0.2) & 1200(325) & PNS & FINE \\
    18 & \cite{Davis2016RestoringNerves} & S1 median & 222(133) & 27(20) & iPNS & USEA \\
    19 & \cite{Davis2016RestoringNerves} & S2 ulnar & 143(76) & 12(11) & iPNS & USEA \\
    20 & \cite{George2020Long-termMuscles} & S5-M first & \text{81.6 (99.6)} & \text{25 (17)} & iPNS & USEA \\
    21 & \cite{George2020Long-termMuscles} & S5-U first & \text{77.5 (101.5)} & \text{31 (31)} & iPNS & USEA \\
    22 & \cite{George2020Long-termMuscles} & S6-M first & \text{67.6 (130.6)} & \text{21 (11)} & iPNS & USEA \\
    23 & \cite{George2020Long-termMuscles} & S6-U first & \text{49.0 (71.4)} & \text{37 (43)} & iPNS & USEA \\
    24 & \cite{George2020Long-termMuscles} & S5-M last & \text{131.3 (163.8)} & \text{60 (40)} & iPNS & USEA \\
    25 & \cite{George2020Long-termMuscles} & S5-U last & \text{178.6 (65.9)} & \text{70 (53)} & iPNS & USEA \\
    26 & \cite{George2020Long-termMuscles} & S6-M last & \text{50.6 (42.8)} & \text{73 (25)} & iPNS & USEA \\
    \bottomrule
    \end{tabular}
    \end{threeparttable}%
    \end{adjustbox}
\end{table}

\subsection{Data Analysis}\label{sec:DataAnalysis}
The power losses at the output are the result of a mismatch between the supply voltage and the channel-specific load voltage. To compare the impact of different voltage scaling strategies, channel-specific voltage requirements need to be calculated. To that purpose, a numerical dataset with \num{100000} entries per subject listed in  \cref{tab:Datasets} was created. The current amplitude and electrode impedance data for these datasets were calculated using three different methods, depending on the available information. For variables reported as mean ± sd, the dataset was assumed to follow a truncated normal distribution with the given parameters. The distribution was truncated at the reported extreme values or at the minimal step size of the parameter, ensuring no negative values were generated. For data provided as a dataset by the authors, the probability density function (PDF) of the variable was estimated using kernel density estimation. The dataset was then filled with values such that the variable followed the estimated PDF. In cases where the data was reported as the median and IQR, the distribution is also estimated to be normal. Although \cite{George2020Long-termMuscles} mentions that the data is not normally distributed, the lack of additional information on the distribution led us to assume a normal distribution as a reasonable estimation. Similar to the mean ± sd data, the dataset followed a truncated normal distribution, with the mean and standard deviation estimated from the median and IQR values, respectively.

In the resulting dataset, each entry received a random value for the current amplitude ($I_{\textrm{th}}$) and electrode impedance ($Z$), following these distributions. For simplification of the calculations, this work assumes the combined impedance of the electrode-tissue interfaces and the tissue, $Z$, to be real (resistive) and equal to the impedance magnitude measured at \SI{1}{\kHz}. The required load voltage at each entry was then calculated using the following equation
\begin{equation} \label{eq:Vload}
    V_{\textrm{load,}i} = I_{\textrm{th,}i} Z_i.
\end{equation}
Using this dataset, the power losses of different voltage scaling strategies were calculated as described in \cref{subsec:Plosses}. The resulting data is available at \cite{datarepository}.

\subsection{Calculating Power Losses}\label{subsec:Plosses}
To calculate the power losses for each scaling strategy, a Monte Carlo sampling method is used. This method involves the following steps:
\begin{enumerate}[label=\arabic*.]
    \item For each subject in the dataset, a subset of $M$ samples is randomly chosen from the dataset described in \cref{sec:DataAnalysis}. The size of $M$ is tailored to the target application.
    \item On each subset, the power losses on each channel are calculated with the methods outlined below.
    \item The efficiencies and power losses of the subset are averaged to obtain the expected average efficiency and power loss per channel for each application. 
    \item The sampling method is repeated for $n_{\rm repeats}$ repetitions on each subject. In this work a repetition rate of $n_{\rm repeats} = 1000$ was used.
\end{enumerate}

For the first step, the size of $M$ needs to be determined for each application. The size of $M$ matters mainly for the calculation of the power losses in case of a global scaling supply, but the subset is applied to the calculation for all methods to ensure fair comparison of the methods. For intracortical visual prostheses, current efforts are aimed at developing systems with more than \num{1000} channels to provide high-resolution visual information that could restore useful vision \cite{Fernandez2020TowardProspects}. In retinal prostheses, it has been estimated that \num{625} channels would be sufficient for useful vision \cite{Margalit2002RetinalBlind}. However, using smaller and more electrodes could improve the field of view and efficacy of the implant \cite{Palanker2005}. In the case of PNS applications, the channel requirements are generally much lower. For intrafascicular interfaces, such as those using the USEA, one or two arrays, each with \num{100} channels, can provide sufficient information for neuroprosthetic applications \cite{George2020Long-termMuscles}. This suggests that a total channel count of approximately \num{200} may be adequate for many tasks. On the other hand, extraneural electrodes, such as the FINE, offer much lower resolution, with individual electrodes typically containing only \num{8} channels. Therefore, a system utilizing two FINE electrodes would have only \num{16} channels in total \cite{Tan2015StabilityAmputees}. Additionally, only a subset of the total channel will be active at the same time. For this analysis, it is assumed that only \SI{20}{\%} of the channels will be active simultaneously. Consequently, in this study, the sample size $M$ is set to 200, 125, 40, and 4 for V1, Retinal, iPNS, and extraneural PNS applications, respectively. Note that the exact value for $M$ is not important, but the order of magnitude is, as will be shown in the results. When repeating this method for a new application, the sample size can be iterated to evaluate the impact of the number of channels on the efficacy of each scaling strategy. 

To calculate the power loss at the output, the load power in the ideal case, when the voltage supply tracks the load voltage accurately for each channel (\cref{fig:IdealSupply}), is used as a reference. In the ideal case, all the power at the output is delivered to the load. Using the current and impedance information of the samples, the load power can be calculated as:
\begin{equation} \label{eq:Pload}
    P_{\textrm{load,}i} = I_{\textrm{th,}i}^2Z_i.
\end{equation}
Subsequently, the efficiency for all non-ideal cases is calculated using:
\begin{equation}
    \eta = \frac{P_{\textrm{load,}i}}{P_{\textrm{load,}i} + P_{\textrm{loss,}i}},
\end{equation}
where $P_{\rm loss}$ are the losses associated with the voltage scaling strategy. 

In the case of a fixed voltage supply (\cref{fig:FixedSupply}), the power losses at each channel can be calculated using:
\begin{equation}
    P_{\textrm{loss,fixed,}i} = (V_{\textrm{fixed}} - V_{\textrm{load,}i})I_{\textrm{th,}i},
\end{equation}
where $V_\textrm{{fixed}}$ is the same for all channels. When designing a system to deliver the stimulation from a fixed voltage supply, different considerations could lead to the choice of $V_\textrm{fixed}$. In this work, $V_\textrm{fixed}$ is based on grouping all data with the same target (in \cref{tab:Datasets}) together. Furthermore, a reasonable design consideration is the trade-off between channel yield and efficiency. Here, channel yield is defined as the percentage of the total number of available channels that can be stimulated. If $V_\textrm{fixed} < \max(V_\textrm{load,target})$, not all channels can be stimulated, resulting in a lower yield, but the overall system efficiency will improve. Whether it can be tolerated to allow for yield \SI{<100}{\%} depends on the application and the design requirements. In this work, a yield of \SI{75}{\%} is chosen for most calculations, unless stated differently. In other words, $V_\textrm{fixed}$ is equal to the third quantile ($Q_3$) of $V_\textrm{load,target}$. Note that a yield of \SI{75}{\%} might be considered low for some applications and that higher yields are desirable. This will favor the more flexible scaling strategies as the variability in load voltages increases. The effect on the yield choice will also be considered in the results section.
In practice, the technology used to design the stimulator circuit will also influence the choice of $V_\textrm{fixed}$ as some values (e.g., \SI{3.3}{\V}, \SI{5}{\V}, \SI{10}{\V}) are common for given technologies. This consideration is not included in the analyses presented in this work, but the method could be repeated if these constraints are known.

The first supply scaling strategy considered in this work is the use of a globally scaled supply voltage (\cref{fig:GlobalScaling}). In this case, the global supply would have to accommodate for the worst-case channel that is being stimulated. The power losses for this strategy are calculated as:
\begin{equation}
    P_{\textrm{loss,global,}i} = (\max(V_{\textrm{load,}j})-V_{\textrm{load,}i})I_{\textrm{th,}i}, j \in M.
\end{equation}
Thus, the maximum load voltage in the subset $M$ is used as the supply voltage for all channels in the sampled subset. 
 
The other supply scaling strategy is the use of a stepped supply voltage (\cref{fig:SteppedSupply}). In this case, $N$ voltage steps are created, and each channel is connected to the nearest available step above the load voltage. Thus, the power losses can be calculated as: 
\begin{equation}
    P_{\textrm{loss,stepped,}i} = (V_{\textrm{step,}i}-V_{\textrm{load,}i})I_{\textrm{th,}i},
\end{equation}
where $V_{\textrm{step,}i}$ is the nearest available voltage rail above $V_{\textrm{load,}i}$.
The available supply rails depend on the numbers of steps chosen. For the results in this work, the voltage rails are calculated using a uniform distribution of the rails between \SI{0}{\V} and $V_{\rm fixed}$. As an example, in the case of $V_{\textrm{fixed}} =$ \SI{5}{\V} and $N =$ \num{4}, the available voltage steps will be \SIlist{1.25;2.5;3.75; 5}{\V}. The methodology is not limited to a uniform distribution and could be repeated using any desired distribution of the voltage rails.

\section{Results}
\subsection{Voltage and load power distributions}\label{subsec:Distributions}
\begin{figure}[ht]
    \centering
    \begin{subfigure}[b]{.49\textwidth}
        \centering
        \includegraphics[width=\textwidth]{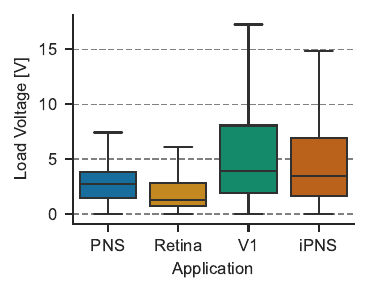}
        \caption{}
        \label{fig:LoadVoltageDistribution}
    \end{subfigure}
    \hfill
    \begin{subfigure}[b]{.49\textwidth}
        \centering
        \includegraphics[width=\textwidth]{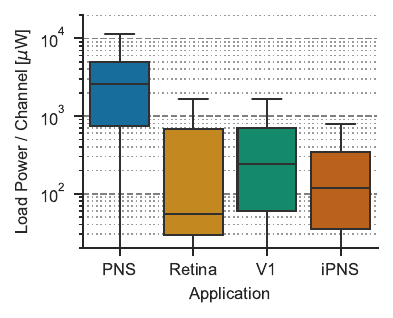}
        \caption{}  
        \label{fig:PloadDistribution}
    \end{subfigure}
    \caption{Load (a) voltage and (b) power per channel distributions grouped by application.}
\end{figure}

The calculated load voltage distributions (\cref{eq:Vload}) are shown in \cref{fig:LoadVoltageDistribution}. The resulting load voltages for the different applications are \medianIQR{3.5}{5.3}{\volt} (median (IQR)), \medianIQR{3.9}{6.1}{\volt}, \medianIQR{1.3}{2.2}{\volt}, and \medianIQR{2.8}{2.4}{\volt} for iPNS, V1, Retina, and PNS, respectively. These results suggest that PNS and Retina stimulation operate at relatively lower voltages compared to iPNS and V1, reflecting varying requirements across applications.

The corresponding load power distributions (\cref{eq:Pload}) are presented in \cref{fig:PloadDistribution}. The resulting values are as follows: \medianIQR{117}{306}{\uW} for iPNS, \medianIQR{243}{637}{\uW} for V1, \medianIQR{55}{656}{\uW} for Retina, and \medianIQR{2.6}{4.2}{\mW} for PNS. These results show that the median load power per channel is highly application-dependent, spanning more than one order of magnitude between Retinal and PNS stimulation. 

\begin{figure}[ht!]
    \centering
    \includegraphics[width=.7\textwidth]{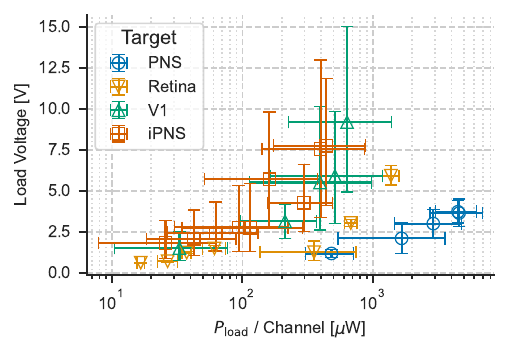}
    \caption{Load voltage and power distributions of the individual datasets. Markers indicate the median value and error bars the IQR.}
    \label{fig:PloadVload}
\end{figure}

To further explore inter-subject differences, the load voltage and power distributions of all subjects are compared in \cref{fig:PloadVload}. This figure highlights the differences between applications. For Retinal stimulation, the load power spans a wide range on the application level, while the range within each subject is small. On the other hand, both iPNS and V1 stimulation show wide ranges both on the application level and on the subject level. 

\begin{figure}[ht!]
    \centering
    \includegraphics[]{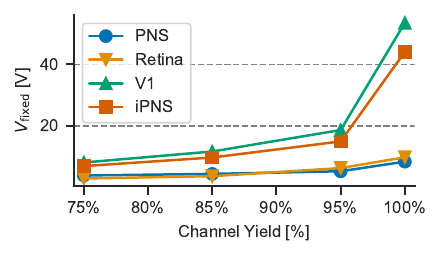}
    \caption{$V_{\rm fixed}$ for different channel yield settings across the different applications.}
    \label{fig:VappvsYield}
\end{figure}

The application-specific voltage supplies for a channel yield of \SI{75}{\%}, used in subsequent power loss calculations, are listed in \cref{tab:TargetVoltages}. The effect of channel yield on the voltage supply is illustrated in \cref{fig:VappvsYield}. As shown there, a yield of \SI{100}{\%} would require a supply voltage of \SI{44}{\V} and \SI{54}{\V} for the applications of iPNS and V1, respectively. Next to the inefficiency that this would cause, it would also require special circuits that can generate and handle such voltage levels.

\begin{table}[ht!]
    \centering
    \caption{Application-specific voltage supply (for a channel yield of \SI{75}{\%}) used for the power loss calculations.}
    \label{tab:TargetVoltages}
    \begin{adjustbox}{max width=\textwidth}
    \begin{threeparttable}[b]
    \sisetup{separate-uncertainty=true, table-number-alignment=center}
    \begin{tabular}{@{}cc@{}}
    \toprule
    Application & $V_\textrm{fixed}$ [V] \\
    \midrule 
    iPNS & 7.0 \\
    V1 & 8.1  \\
    Retina & 2.9  \\
    PNS & 3.9 \\ 
    \bottomrule
    \end{tabular}
    \end{threeparttable}%
    \end{adjustbox}
\end{table}

\subsection{Losses with fixed voltage supply}\label{subsec:FixedLosses}
\begin{figure}[ht!]
    \centering
    \includegraphics[]{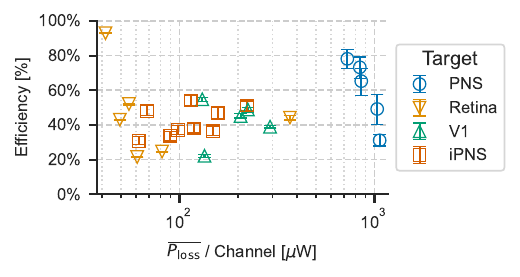}
    \caption{Average power loss per channel (Median \& IQR) and the corresponding efficiencies for a conventional fixed voltage supply ($V_{\rm fixed}$). Each marker corresponds to one subject.}
    \label{fig:scatterPlossEff_Vapp}
\end{figure}

The power losses per channel for a conventional fixed voltage supply are shown in \cref{fig:scatterPlossEff_Vapp}, where the markers denote the different subjects in \cref{tab:Datasets}. Most subjects exhibit efficiencies below \SI{60}{\%}, with power losses typically in the range of \SI{100}{\uW} per channel. However, PNS subjects display higher efficiencies despite experiencing greater power losses in the order of \SI{1}{\mW} per channel. This emphasizes that even though efficiencies may be high, it could still be worth improving to save significant power. Except for the PNS subjects, the plot shows small variations within each subject, which can be attributed to two factors. In the Retina subjects, the spread in load power and load voltage within each subject is small, as shown in \cref{fig:PloadVload}. For the V1 and iPNS subjects, the small variation is likely an effect of the sample size for resampling since the resampling and averaging filters out extreme values. This shows that, within a subject, the losses for a fixed voltage supply would be predictable and constant for different subsets of channels.

\subsection{Losses for global supply scaling}
\begin{figure}[ht]
    \centering
    \begin{subfigure}[b]{\textwidth}
        \centering
        \includegraphics[width=.8\textwidth]{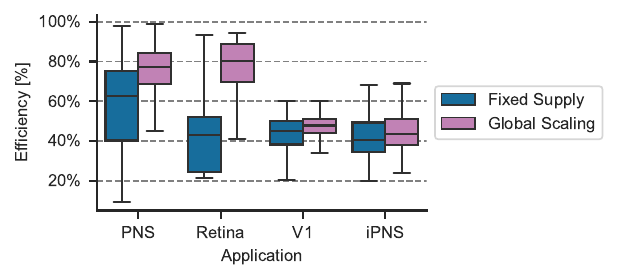}
        \caption{}
        \label{fig:Eff_global}
   \end{subfigure}
   \begin{subfigure}[b]{\textwidth}
        \centering
        \includegraphics[width=.8\textwidth]{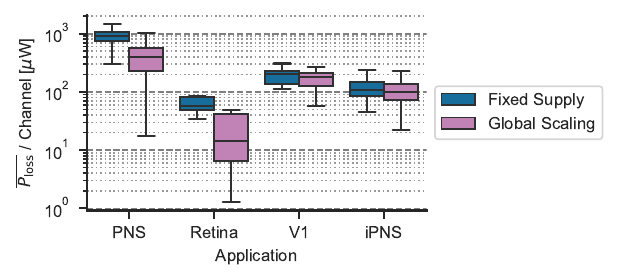}
        \caption{}
        \label{fig:Ploss_global}
   \end{subfigure}
    \caption{The effect of a global supply scaling strategy on the (a) efficiency and (b) power loss per channel for the different applications. Outliers are not shown in the plot.}
    \label{fig:Global}
\end{figure}

The efficiencies and power loss per channel for global supply scaling are compared to fixed voltage supplies in \cref{fig:Global}. In applications with high channel counts and wide spread in load conditions (iPNS and V1) the efficiency is negligible due to the fact that the supply has to accommodate the worst case channel. However, in low-channel-count applications like PNS, the approach can reduce power losses substantially. For PNS, the median power loss per channel is reduced from \SI{914}{\uW} to \SI{404}{\uW}, while the median efficiency is increased from \SI{62.9}{\%} to \SI{77.3}{\%}. Furthermore, if the variability in load conditions within subjects is small, as is the case for the Retina data, global supply scaling also leads to significant improvements. In the Retina data, the median power loss per channel is reduced from \SI{58}{\uW} to \SI{14}{\uW}, while the median efficiency is increased from \SI{43.1}{\%} to \SI{80.2}{\%}. These results show that global scaling is most effective when the channel count is low and the variability in voltage requirements within subjects is small.

\subsection{Losses for a stepped supply}
\begin{figure}[ht]
    \centering
    \begin{subfigure}[b]{\textwidth}
        \centering
        \includegraphics[]{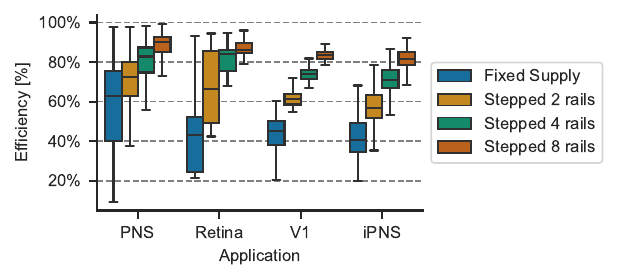}
        \caption{}
        \label{fig:Eff_stepped_Vapp}
    \end{subfigure}   
    \begin{subfigure}[b]{\textwidth}
        \centering
        \includegraphics[]{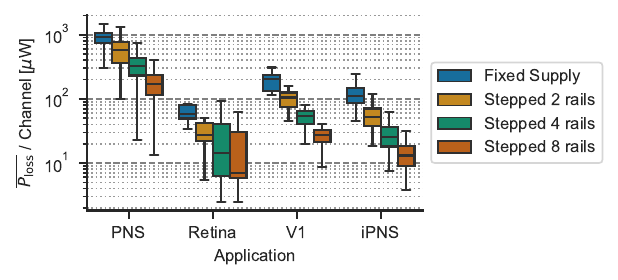}
        \caption{}
        \label{fig:Ploss_stepped_Vapp}
    \end{subfigure}
    \caption{Effect of a stepped-voltage supply strategy from an application-specific supply on the (a) efficiency and (b) power loss per channel in the different applications. Outliers are not shown in the plot.}
    \label{fig:Stepped_Vapp}
\end{figure}

For the stepped supply strategy, the power loss reduction and efficiency improvements are calculated for uniformly distributed supplies of 1 (fixed), 2, 4, and 8 voltage rails. This strategy demonstrates efficiency improvements across all applications (\cref{fig:Stepped_Vapp}), with efficiencies exceeding \SI{81}{\%} when using eight voltage rails. Compared to the fixed voltage supply, this yields an increase in efficiency of \SI{43}{\%} (PNS) to \SI{100}{\%} (iPNS and Retina). However, the incremental benefit of adding more rails diminishes with each step. Nonetheless, the flexibility to tune each channel specifically makes this strategy broadly applicable. 

\subsection{Comparison of scaling strategies}
\begin{figure}[ht]
    \centering
    \begin{subfigure}[b]{\textwidth}
        \centering
        \includegraphics[]{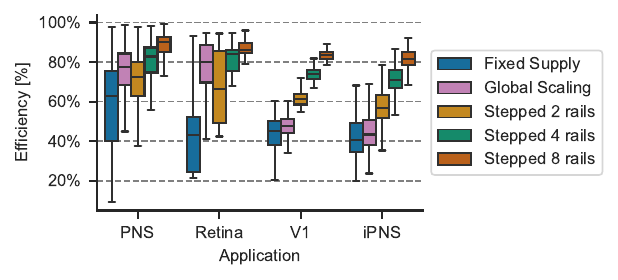}
        \caption{}
        \label{fig:Eff_compare}
    \end{subfigure}   
    \begin{subfigure}[b]{\textwidth}
        \centering
        \includegraphics[]{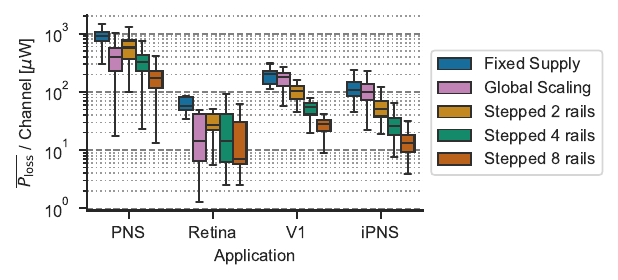}
        \caption{}
        \label{fig:Ploss_compare}
    \end{subfigure}
    \caption{Comparison of the impact of the scaling strategies on the (a) efficiency and (b) power loss per channel in the different applications. Outliers are not shown in the plot.}
    \label{fig:Compare}
\end{figure}

The comparative performance of all strategies is summarized in \cref{fig:Compare}. Furthermore, the normalized efficiencies and power losses are listed in \cref{tab:Normalization_comparison}, where each value is normalized to the Fixed Voltage configuration for each respective application. It is shown that stepped supplies with \num{4} and \num{8} rails outperform the global scaling across all applications, although for Retina stimulation the performance of global scaling and stepped 4 rails is comparable. Furthermore, in some cases, one strategy outperforms the others, but other factors, such as design complexity and circuit losses, may still favor another strategy. For example, in Retina data, the relative improvement from 4 rails to 8 rails is \SI{2}{\%} corresponding to a reduction of \SI{7}{\uW} per channel, which is likely not worth the extra circuits. Considerations regarding circuit implementations will be discussed in \cref{sec:CircuitConsiderations}. 
\begin{table}[ht!]
\centering
\caption{Comparison of normalized efficiency ($\eta$) and normalized power loss ($P_{\rm loss}$) across different applications and scaling strategies. All values are normalized to the Fixed Supply configuration for each respective application.}
\label{tab:Normalization_comparison}
\resizebox{\textwidth}{!}{%
\begin{tabular}{@{}l S[table-format=1.2] S[table-format=1.2] S[table-format=1.2] S[table-format=1.2] S[table-format=1.2] S[table-format=1.2] S[table-format=1.2] S[table-format=1.2] @{}}
\toprule
\multirow{2}{*}{Application} & \multicolumn{2}{c}{Retina} & \multicolumn{2}{c}{V1} & \multicolumn{2}{c}{PNS} & \multicolumn{2}{c}{iPNS} \\ 
\cmidrule(lr){2-3} \cmidrule(lr){4-5} \cmidrule(lr){6-7} \cmidrule(lr){8-9}
                 & \textbf{$\eta /\eta_{\rm ref}$} & \textbf{\(P_{\rm loss}/P_{\rm ref}\)} & \textbf{$\eta /\eta_{\rm ref}$} & \textbf{\(P_{\rm loss}/P_{\rm ref}\)} & \textbf{$\eta /\eta_{\rm ref}$} & \textbf{\(P_{\rm loss}/P_{\rm ref}\)} & \textbf{$\eta /\eta_{\rm ref}$} & \textbf{\(P_{\rm loss}/P_{\rm ref}\)} \\ \midrule
Fixed Supply     & 1.00  & 1.00  & 1.00  & 1.00  & 1.00  & 1.00  & 1.00  & 1.00  \\
Global Scaling   & 1.86  & 0.25  & 1.06  & 0.89  & 1.23  & 0.44  & 1.07  & 0.90  \\
Stepped 2 rails  & 1.54  & 0.47  & 1.36  & 0.51  & 1.15  & 0.63  & 1.40  & 0.47  \\
Stepped 4 rails  & 1.95  & 0.25  & 1.64  & 0.27  & 1.32  & 0.36  & 1.74 & 0.24 \\
Stepped 8 rails  & 2.00  & 0.12  & 1.85  & 0.14  & 1.43  & 0.19  & 2.00  & 0.12  \\ \bottomrule
\end{tabular}%
}
\end{table}

Even though the evaluated strategies show substantial improvements in efficiency, there is still room for further reduction of the power losses. For each application, the total power loss can be calculated by multiplying the number of channels in that application with the channel losses presented in \cref{fig:Ploss_compare}. For the best scaling strategy in each application, this results in a total system power loss of \medianIQR{525}{366}{\uW}, \medianIQR{5.5}{2.4}{\mW}, \medianIQR{879}{3016}{\uW}, \medianIQR{683}{466}{\uW} for iPNS, V1, Retina, and PNS, respectively. 

\subsection{Effect of Channel Yield}
\begin{figure}[ht!]
    \centering
    \includegraphics[width=\textwidth]{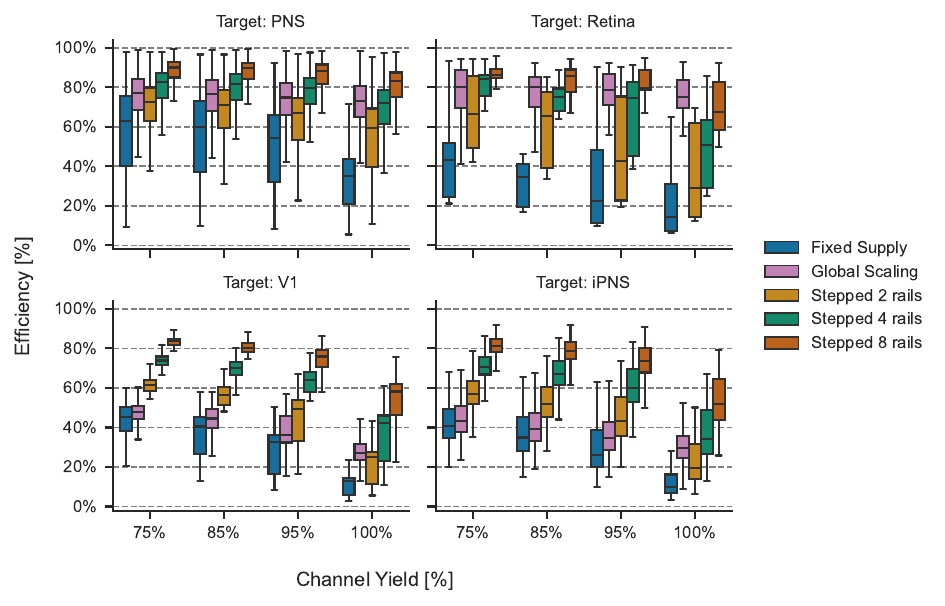}
    \caption{Effect of channel yield on the efficiency across different applications and scaling strategies.}
    \label{fig:Eff_vs_Yield}
\end{figure}

\begin{figure}[ht!]
    \centering
    \includegraphics[width=\textwidth]{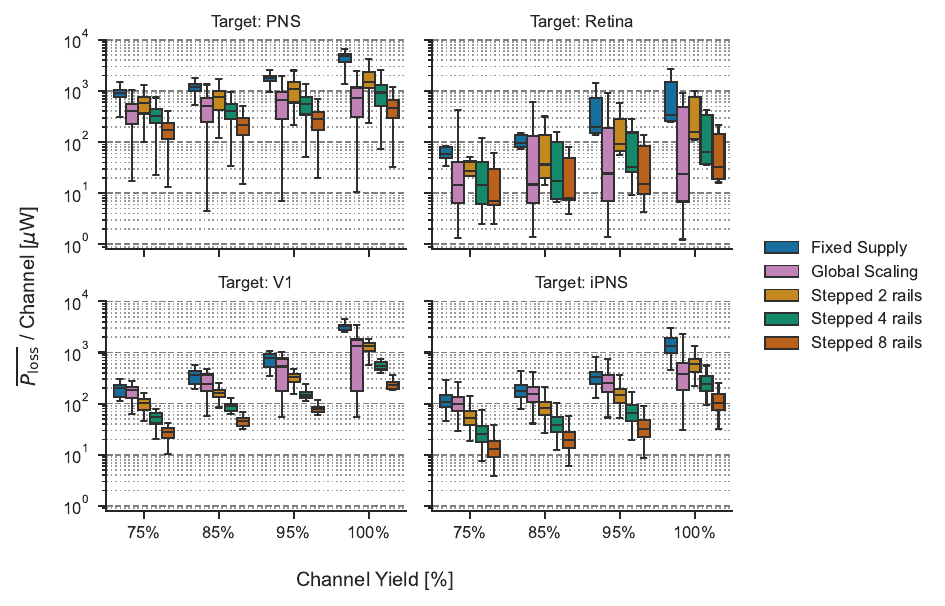}
    \caption{Effect of channel yield on the power losses across different applications and scaling strategies.}
    \label{fig:Ploss_vs_Yield}
\end{figure}

All results so far were calculated with a channel yield of \SI{75}{\%}. The effect of the channel yield on the efficiency and power losses is shown in \cref{fig:Eff_vs_Yield,fig:Ploss_vs_Yield}. First of all, increasing the yield diminishes the efficiency and increases the power losses for all strategies across all applications. However, the global scaling strategy is more robust to changes in the yield. At several conditions the global scaling becomes more efficient than the stepped 4 rail supply, and for a yield of \SI{100}{\%} it is the most efficient strategy for the Retina data. One possible improvement for the stepped voltage supplies would be to use non-uniformly distributed voltage rails. Furthermore, as mentioned in \cref{subsec:Distributions}, achieving a higher yield also increases the circuit complexity due to the increased voltage supply. Therefore, the trade-off between channel yield and power efficiency will also be impacted by practical limitations of the implementation. The proposed method allows to evaluate each condition to guide the design process.  
\section{Discussion}
\subsection{Circuit Design Considerations}\label{sec:CircuitConsiderations}
The voltage supply levels in this work are based on the available datasets, without taking into account the implementation of these voltage rails. In reality, most systems will be designed adhering to common voltage levels depending on the technology used for the circuit design. The first use of the proposed method could be to get an estimation of the channel yield for a given voltage supply level to decide is that is sufficient for the application. Furthermore, the method can be used to decide whether implementing more complex supply strategies is worthwhile.

In the analyzed applications, global scaling showed limited benefit over a fixed supply voltage in scenarios with high channel counts and a wide spread of load conditions. However, in applications with fewer channels or when within-subject load variation is low, global scaling can significantly improve efficiency\textemdash up to \SI{86}{\%} in the Retina dataset. Several designs in the literature adopt a global scaling supply \cite{Ortmanns2007,Noorsal2012,Lee2013}. The added complexity of these implementations is a voltage compliance monitor at every channel to determine what the supply voltage should be. 

A stepped voltage supply with four or more rails generally outperformed global scaling, especially when intrasubject variability was high (e.g., in V1 and iPNS). In principle, increasing the number of voltage rails reduces losses, but it also raises complexity at both the system and channel levels. At the system level, increasing the number of voltage rails in the multi-output supply increases its complexity. Each voltage rail requires a storage capacitor to make it possible to deliver power, which can increase the area requirements.
Furthermore, all voltage rails need to be distributed towards all channels. In high-channel-count applications, this can become a bottleneck for increasing the number of rails. At the channel level, more rails require a larger selector circuit. Since the area requirements are highly technology-dependent, it is difficult to estimate the cost of each implementation. However, once the technology parameters are known, the presented analysis can serve to make a trade-off between efficiency, area, and complexity. 

Finally, for each scaling strategy, additional power losses should be considered. Additional losses could come from reduced rectifier efficiency, extra conversion steps, and additional control and compliance circuitry. In this respect, the power per channel for the application is an important weighting factor. When the load power per channel is low, overhead losses in the channel can quickly diminish the efficiency improvements gained by the scaling strategy. Sharing resources among multiple channels helps to reduce the impact on the efficiency. When the power per channel is high, the design requirements for the scaling circuitry are easier in terms of power consumption since overhead losses are less important. 

\subsection{Temporal changes}
The datasets used in this study all represent a static set of parameters. However, impedance and stimulation thresholds are known to change over time \cite{Davis2012,Chen2023ChronicMonkeys,George2020Long-termMuscles}. Therefore, the voltage requirements of the channels and resulting losses will change accordingly. Flexibility in the voltage supply helps to accommodate changes and reduce power losses over time. In the long term, this will lead to the best power efficiency during the lifetime of the implant.

\subsection{Limitations of current work}
The analysis presented here is based on previously published data. Here, we reflect on the limitations of the assumptions necessary to perform the analysis. 

First, for most datasets, the distribution of the parameters was assumed to be (truncated) Gaussian. This assumption was made due to a lack of information; in reality, the distributions could have been different. As described in \cref{sec:DataAnalysis}, an extended analysis was performed when more information regarding the data distribution was available. 

Furthermore, the impedance and current data are assumed to be uncorrelated. While \cite{DeBalthasar2008FactorsProstheses} found a negative correlation between the two parameters, the other studies did not report the correlation. Variations in electrode impedance are caused by many factors, which might change the correlation between impedance and threshold values. In \cite{DeBalthasar2008FactorsProstheses}, the critical factor influencing impedance and threshold was the distance between the electrode and the retina. If the correlation of these parameters is known for a specific application, it could be added to the generation of the dataset to evaluate its effects.

Additionally, the size of the electrodes used in the retina data is relatively big. The development trend in retinal implants is to reduce the size of the electrodes to bring them closer to the retina and achieve higher electrode count and density \cite{Ho2019CharacteristicsArrays,Palanker2020PhotovoltaicDegeneration}. However, no data could be found on human subjects for the smaller electrodes, where both impedance measurements and perception thresholds were reported. Generally speaking, reducing the electrode size will increase the impedance, and bringing the electrodes closer to the cells will reduce stimulation thresholds \cite{DeBalthasar2008FactorsProstheses}.

Last, the electrode impedance is more complex than the \SI{1}{\kHz} value used in this work to calculate the losses. A more realistic model includes the capacitive effects of the electrode-tissue interface (ETI). This capacitive component affects the load voltage and will change the load power. The extent of this effect depends on the ratio between the resistive and capacitive components of the load impedance, as well as the delivered charge \cite{Varkevisser2022EnergyLosses}. Therefore, it depends on the type of electrodes used. Ideally, the capacitance of stimulation electrodes should be large to prevent depolarization of the ETI, which can lead to non-reversible charge transfer \cite{Merrill2005}. Specifically, in microelectrodes, the resistive component is typically dominant, and the effect of the capacitance on the power calculations will be minimal.  

\section{Conclusion}
This work introduces a novel methodology for analyzing power losses in multichannel electrical stimulation systems, integrating both electrophysiological and electronic considerations. Traditional stimulator systems are often designed for fixed load conditions, overlooking the significant impact of inter-channel variability in electrode impedance and current thresholds. By incorporating these variabilities into the analysis, the proposed method enables more elaborate assessments of the power efficiency across various scaling strategies and applications. Furthermore, the method serves as a tool for guiding the design of new systems, providing insights into which scaling strategy offers the best performance under specific conditions.

Applied to experimental data from multiple multichannel systems, the methodology reveals that a stepped voltage supply with 8 voltage rails can boost efficiency by \SIrange{43}{100}{\%}, proving to be most effective for high-channel-count applications with significant inter-channel variation. Conversely, global voltage scaling emerged as a viable option for applications with fewer channels or minimal inter-channel variability. These findings underscore the critical role of application-specific parameters, such as channel count and load variance, in selecting the most suitable voltage scaling approach. 

Furthermore, while advanced supply strategies can substantially reduce power losses, they invariably add complexity at both the system and channel levels. The specific cost–benefit trade-offs depend on the underlying technology and target application, making generalization challenging. Nonetheless, when specific design targets and technology limitations are known, the proposed methodology can guide the design trade-offs to choose the best approach.

Finally, the calculated total system power losses indicate that there is still room for improvements in more advanced methods to increase power efficiency even further. By developing novel systems that support voltage scaling techniques, power efficiency can be enhanced, allowing for increasing the number of stimulation channels in next-generation, large-scale neural interfaces.

\ack
We want to thank the authors of the original datasets, specifically Xing Chen and Jacob George, for sharing additional data and insights. This research was supported by NWO, the Dutch Research Council, under project number 17619 ‘INTENSE.’

\end{document}